\documentclass[
 reprint,
 prb,superscriptaddress
]{revtex4-1}

\usepackage{bm}%
\usepackage[colorlinks=true,linkcolor=blue]{hyperref}%
\expandafter\ifx\csname package@font\endcsname\relax\else
 \expandafter\expandafter
 \expandafter\usepackage
 \expandafter\expandafter
 \expandafter{\csname package@font\endcsname}%
\fi
\usepackage{graphicx}
\usepackage{dcolumn}

\begin{document}

\preprint{APS/123-QED}

\title{Metallic-insulator phase transitions in the extended Harper model}

\author{Jo\~ao Chakrian}
\affiliation{Departamento de F\'{\i}sica, Universidade Federal Rural de Pernambuco, 52171-900, Recife, PE, Brazil}

\author{Marcelo L. Lyra}
\email{marcelo@if.ufal.br}
\affiliation{Instituto de F\'isica, Universidade Federal de Alagoas, 57072-970, Macei\'o - AL, Brazil}

\author{Jonas R. F. Lima}
\email{jonas.de-lima@uni-konstanz.de}
\affiliation{Departamento de F\'{\i}sica, Universidade Federal Rural de Pernambuco, 52171-900, Recife, PE, Brazil}
\affiliation{Department of Physics, University of Konstanz, 78457 Konstanz, Germany}

\date{\today}

\begin{abstract}

In this work we investigate the transport properties of non-relativistic quantum particles on incommensurate multilayered structures with the thicknesses $w_n$ of the layers following an extended Harper model given by $w_n = w_0 |\cos(\pi a n^{\nu})|$. For the normal incidence case, which means an one-dimensional system, we obtained that for a specific range of energy, it is possible to see a metallic-insulator transition with the exponent $\nu$. A metallic phase is supported for $\nu<1$. We also obtained that for the specific value $\nu=1$ there is an alternation between metallic and insulator phases as we change the disorder strength $w_0$. When we integrate out all incidence angles, which means a two-dimensional system, the metallic-insulator transition can be seen for much larger range of energy compared to the normal incidence case.

\end{abstract}


\maketitle

\section{Introduction}

One-dimensional (1D) models play an important role in the study of the  electronic transport of solids. This class of models captures the main relevant features related to the emergence of energy bands, the nature of the electronic eigenstates, the metallic or insulator character, the influence of defects and interfaces, etc \cite{mermin,oxford}. 

The characteristics of the interaction potential of the electron with the underlying lattice has a strong influence of the resulting transport properties. A prototype model accounts for a nearest-neighbors tight-binding description for non-interacting electrons in a 1D single-band system. The quantum transport is assumed to be ruled by a discrete set of Schrodinger equations
\begin{equation}
t(u_{n+1} + u_{n-1}) + V_n u_n = Eu_n
\end{equation}
where $t$ is the nearest-neighbors hopping integral term, $V_n$  s the on-site potential and the $n$-th site of a linear chain. The electronic eigenstate of energy $E$ is assumed to be decomposed as $|\Psi\rangle = \sum_n u_n|n\rangle$, considering the basis of local atomic orbitals. A periodic potential $V_n$ give rises to Bloch-like eigenstates which are fully delocalized along the chain, leading to a conducting regime when the band is partially filled\cite{mermin,oxford}. A fully uncorrelated disordered potential leads to the exponential Anderson localization of all eigenstates, resulting in an insulator character irrespective to the band filling\cite{anderson,lee}. Special correlations in the disorder distribution can support extended states. Dimer-like correlations can induce the emergence of resonant delocalized states at specific energies\cite{dimer1,dimer2}. On the other hand, a band of extended states can survive for long-ranged correlated disorder, with mobility edges separating extended and localized states\cite{lyra,izrailev}. 

Determistic one-dimensional potentials which are incomensurate with the underlying lattice have been studied during the last decades in the context of electronic transport because their statistical features interpolates between those of periodic and random potentials\cite{sokoloff}. In the Harper model, with the potential given by $V_n=V_0\cos{(2\pi\alpha n)}$ and $\alpha$ irrational, all states are extended for $V_0<2t$ and localized for $V_0>t$. In the special case of $V_0=2t$ all states become critical\cite{aubry}. An extension of the Harper model considers a incomensurate potentials with varying period on the form\cite{fishman,sarma1,sarma2}
\begin{equation}
 V_n=V_0\cos{(2\pi\alpha n^{\nu})},
 \label{hmodel}
\end{equation}
where the exponent $\nu$ governs how fast the potential period varies. For $\nu \geq 2$ (fastly varying period) the potential becomes statistically equivalent to a uncorrelated disordered potential and all states are exponentially localized. For $\nu <1$ and $\alpha$ irrational (slowly varying incommensurate potential), there are mobility edges separating extended and localized states at $E= \pm(2t-V_0)$ for $V_0<2t$. In the intermediate regime of $1<\nu<2$, it has been argued that all states are localized except at the band center where the localization length seems to diverge very slowly\cite{fishman,sarma1,sarma2,thouless}. 

The above class of incommensurate potentials has been used in the literature to investigate a diversity of aspects including physical realization using cold atoms trapped in optical lattices\cite{luschen,schreiber,roati}, the influence of interactions and external fields\cite{eilmes,peixoto,morales,ray,malla,patra,saha,settino,cookmeyer,sarkar,sajid,yoo}, nonhermiticity\cite{jiang} as well as quantum information characteristics\cite{gong,messias}. In particular, it has been used to model a multilayered photonic structure with the average transmission of harmonic waves being shown to reach a minimum at a specific value of $\nu <1$\cite{nascimento}. However, a detailed study of quantum transport in slowly varying incommensurate multilayered structure is still missing.

In the present work, we provide a systematic study of the quantum transmission characteristics of incommensurate multilayered structures presenting a slowly varying modulation of the sequence of layer thicknesses. We provide a comparative analysis of the transmission spectrum as a function of the energy and incidence angle of the incoming particle beam. The behaviour of the transmission when one increases the number of regions of the system determines a metallic or insulator phase. We obtain here novel metallic-insulator phase transitions in the system when we change the value of the exponent $\nu$ and also the value of the disorder strength $w_0$. These phase transitions will depend on the energy of the incoming particle beam. 

The work in organized as follows. In Sec. II we solve the Schr\"odinger equation for a quantum particle in the presence of a sequence of potential barriers, where the width of the regions in this superlattice follow the extended Harper model. Using the transfer matrix method we obtain the transmission properties of the system. In Sec. III we discuss the results, where we analyze metallic-insulator phase transitions in the system. The work in concluded in Sec. IV.

\section{Model and Structure}

\begin{figure}[h]
\includegraphics[width=\linewidth]{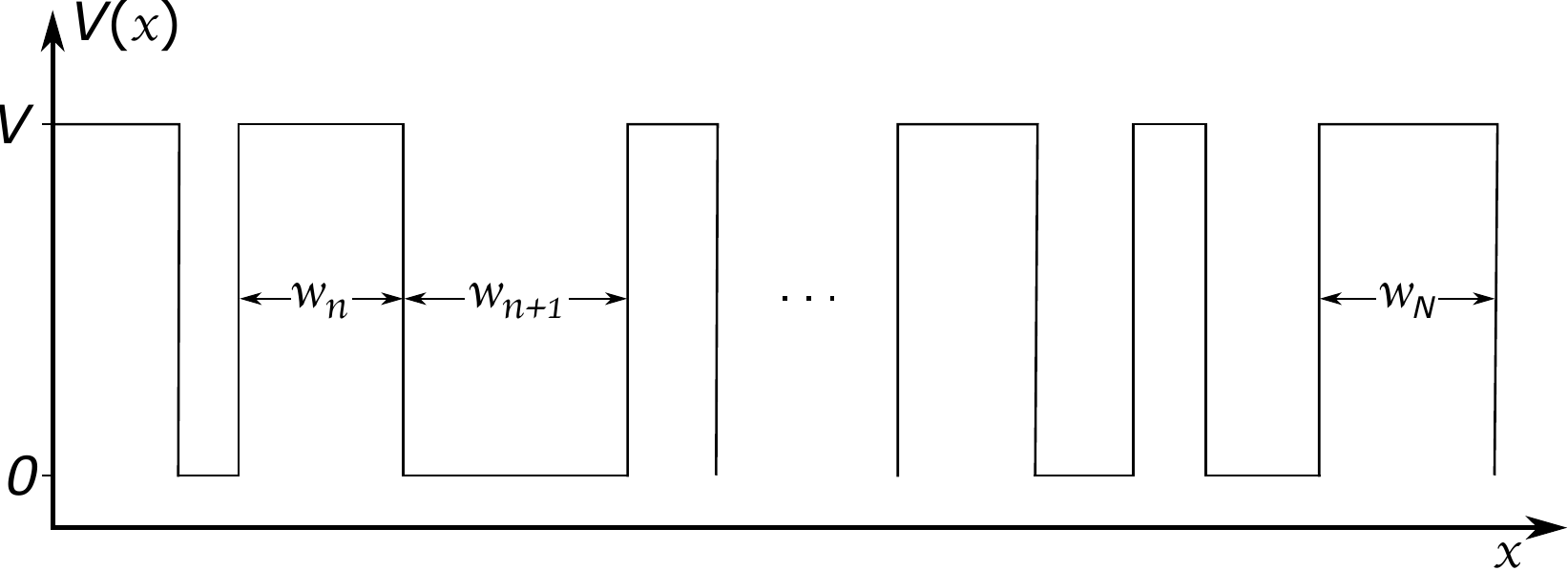}
\caption{\small Schematic diagram of the superlattice structure.} \label{sp}
\end{figure}

We consider here a superlattice composed by a sequence of square potential barriers in the $x$ direction whose potential in each region alternates between $V(x)= V$ and  $V(x) = 0$. The width of each region in the superlattice is modulated following the extended Harper model, which means that the width of the $n$'th region is given by
\begin{equation}
w_n = w_0 |\cos(\pi a n^{\nu})|,
\label{w}
\end{equation}
where $a$ is an irrational number. Here, we are using $a = (\sqrt{5}-1)/2$, the inverse golden ratio. In all the results obtained here, we consider $V = 50$~meV and $w_0 = 1.5$~nm. A schematic diagram of the superlattice can be seen in Fig. \ref{sp}.

The Schr\"odinger equation for non-relativistic quantum particles in two-dimensions under one-dimensional potential barriers is given by
\begin{equation}
    (\partial^2_x + \partial_y^2) \psi(x,y) = -\frac{2m}{\hbar^2}[E-V(x)]\psi(x,y).
   \label{seq}
\end{equation}
The momentum conservation in $y$ direction allows us to write $\psi(x,y)=\psi(x)e^{ik_yy}$. So, Eq. (\ref{seq}) becomes
\begin{equation}
      \partial^2_x \psi(x) = -\frac{2m}{\hbar^2}\left(E-V(x)-\frac{\hbar^2k_y^2}{2m}\right)\psi(x,y).
      \label{scheqx}
\end{equation}
Inside the $j$th region of the superlattice, $V(x)$ is constant. So, the solution of this equation is given by
\begin{equation}
    \psi(x) = \left\{  \begin{array}{cc}
        Ae^{ik_xx} + Be^{-ik_xx}, & \text{regions with V(x) = 0} \\
        A^{\prime}e^{ik_x^{\prime}x} + B^{\prime}e^{-ik_x^{\prime}x},  & \text{regions with V(x) = V},
    \end{array}\right. 
\end{equation}
where
\begin{equation}
    k_x = \frac{\sqrt{2mE-\hbar^2k_y^2}}{\hbar}
    \label{k}
\end{equation}
and
\begin{equation}
    k_x^{\prime} = \frac{\sqrt{2m(E-V)-\hbar^2k_y^2}}{\hbar}.
    \label{kp}
\end{equation}

Considering the continuity of $\psi(x)$ and its first derivative at the interfaces of all regions in a superlattice with $N$ regions, it is possible to write a transfer matrix that connects the amplitude of the electronic waves in the incidence region with the wave in the exit region \cite{soto}, which gives us
\begin{equation}
    \left(\begin{array}{c}
        A   \\
        B  
    \end{array} \right) = M 
     \left(\begin{array}{c}
        C   \\
        0  
    \end{array} \right),
    \label{tmeq}
\end{equation}
where $A(B)(C)$ is the amplitude of the incident (reflected)(transmitted) electronic wave and
\begin{equation}
    M = I_1P_1I_2P_2I_1P_3I_2P_4...I_1P_NI_2.
\end{equation}
Here, we have that
\begin{equation}
    I_i =\frac{1}{t_i} \left(\begin{array}{cc}
       1  & r_i \\
       r_i  & 1
    \end{array} \right),
\end{equation}
with $i=1,2$ and $r_1=(k_x-k_x^{\prime})/(k_x+k_x^{\prime})$, $r_2=(k_x^{\prime}-k_x)/(k_x+k_x^{\prime})$, $t_1=2k_x/(k_x+k_x^{\prime})$ and $t_2=2k_x^{\prime}/(k_x+k_x^{\prime})$. We also have that
\begin{equation}
    P_j = \left(\begin{array}{cc}
       e^{ik_x^jw_j}  & 0 \\
       0  & e^{-ik_x^jw_j}

    \end{array} \right).
\end{equation}

From Eq. (\ref{tmeq}) we can obtain directly the transmittance of the system, which is given by
\begin{equation}
    T = \frac{|C|^2}{|A|^2} = \frac{1}{|M_{1,1}|^2}.
\end{equation}

It is important to mention that in order to obtain the transmission as a function of the incidence angle $\theta_0$, one has to write
\begin{equation}
    k_y = \sin \theta_0 \frac{\sqrt{2mE}}{\hbar}.
\end{equation}

The total conductance of the system at zero temperature can be obtained via the Landauer-B\"uttiker formula, given by
\begin{equation}
G = G_0 \int_{-\pi/2}^{\pi/2} T \cos \theta_0 d\theta_0, 
\end{equation}
where $G_0 = 2e^2 E L_y /(\pi \hbar)$ \cite{conduc}. $L_y$ is the sample size in the $y$ direction.

\section{Results and Discussions}

\subsection{Normal incidence case}

\begin{figure}[h]
\includegraphics[width=\linewidth]{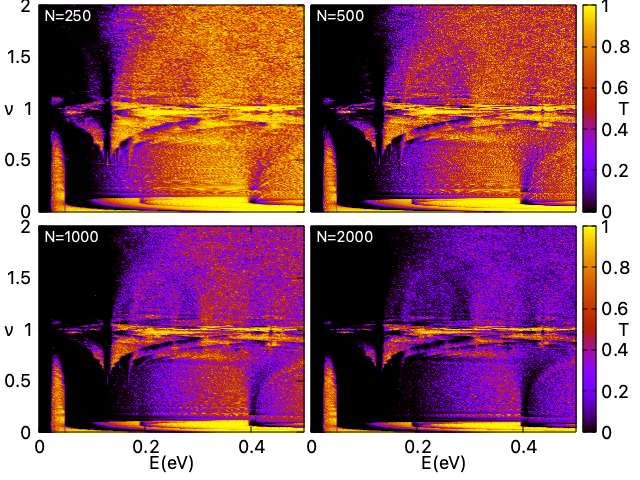}
\caption{\small Contour plot of the transmittance as a function of $E$ and $\nu$ for $\theta_0=0$ and $N=250$, $N=500$, $N=1000$ and $N=2000$.} \label{T1}
\end{figure}

We first consider the contour plot of the transmittance as a function of $\nu$ and $E$ for $\theta_0 = 0$. It can be seen in Fig. \ref{T1}, where we considered $N=250, 500$, $1000$ and $2000$. Since we have only the normal incidence, it represents the one-dimensional version of our problem.

\begin{figure}[b]
\includegraphics[width=\linewidth]{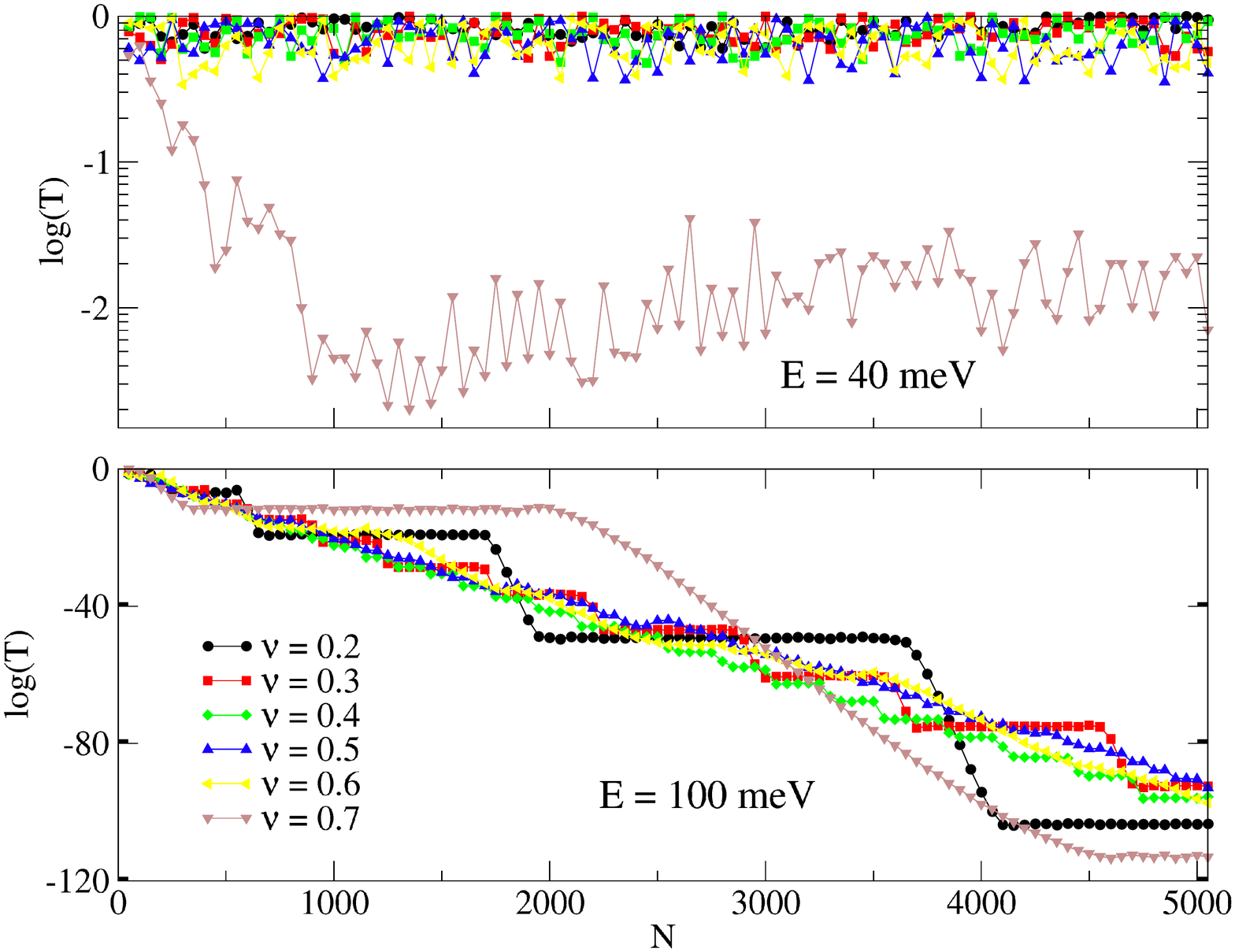}
\caption{\small The transmittance as a function of $N$ for different values of $\nu$.} \label{TxNsmall}
\end{figure}

For $\nu=0$, we have a periodic superlattice, which gives rise to a miniband and minigap structure. As the value of $\nu$ increases but remains small, the minibands and minigaps are shifted down in energy. For $\nu>0.15$ we can not see the minibands anymore. In this case, the transmittance increases with the energy. However, there is a transmission band for $\nu<1$ around $E=0.04$~eV, which suggests the existence of a metallic phase in this region. We can also identify a greater transmission around $\nu=1$. In what follows, we analyze all these behaviours in details. 

 \subsubsection{Small values of $\nu$}

Let us first consider the case of small values of $\nu$. The transmittance as a function of $N$ for different values of $\nu$ is plotted in Fig. \ref{TxNsmall} for $E=40$~meV (top panel) and $E=100$~meV (bottom panel). We can see that, depending on the value of $\nu$, the transmittance decays exponentially or remains constant as the number of barriers increases, which represents an insulator and a metallic phase, respectively.

\begin{figure}[h]
\includegraphics[width=\linewidth]{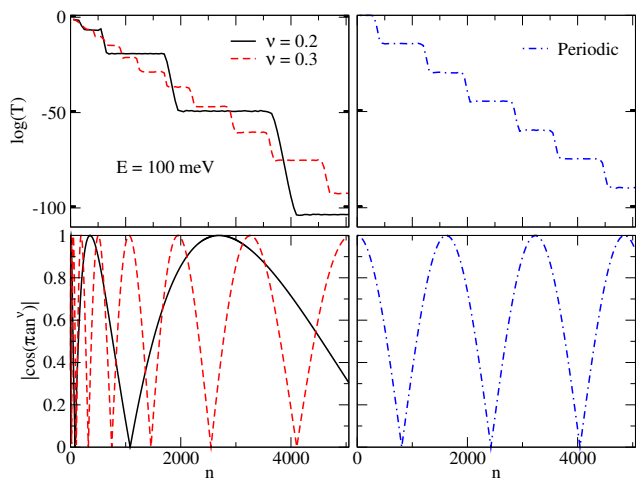}
\caption{\small The transmittance and the cosine that modulates the widths of the regions in the superlattice as a function of $n$ for $\nu=0.2, 0.3$ (left panels) and for a periodic modulation of the widths (right panels). For the periodic case, the argument of the cosine is $\pi anc$, where $c$ is an arbitrary constant.} \label{step}
\end{figure}

In order to understand the metallic and insulator phases revealed in Fig. \ref{TxNsmall}, we consider in Fig. \ref{step} the transmittance for $E=100$~meV and the absolute value of $\cos(\pi a n^{\nu})$ for $\nu$ equal to 0.2 and 0.3 (left panels). One can see that, for small values of $\nu$, the value of the cosine that modulates the widths of the regions in the superlattice changes slowly with $n$. Also, one can note that for a very specific range of values of the cosine, which means specific values of the widths, the transmittance decays, while it remains constant for other values of the cosine, which creates the steps in the transmittance. Step-wise transmissions are focused on system sizes for which the layer widths at the output edge are minimal or maxima. For the sake of comparison, we also consider a periodic modulation of the widths in the Fig. \ref{step} (right panels), and the steps in the transmittance are also observed. 

\begin{figure}[t]
\includegraphics[width=\linewidth]{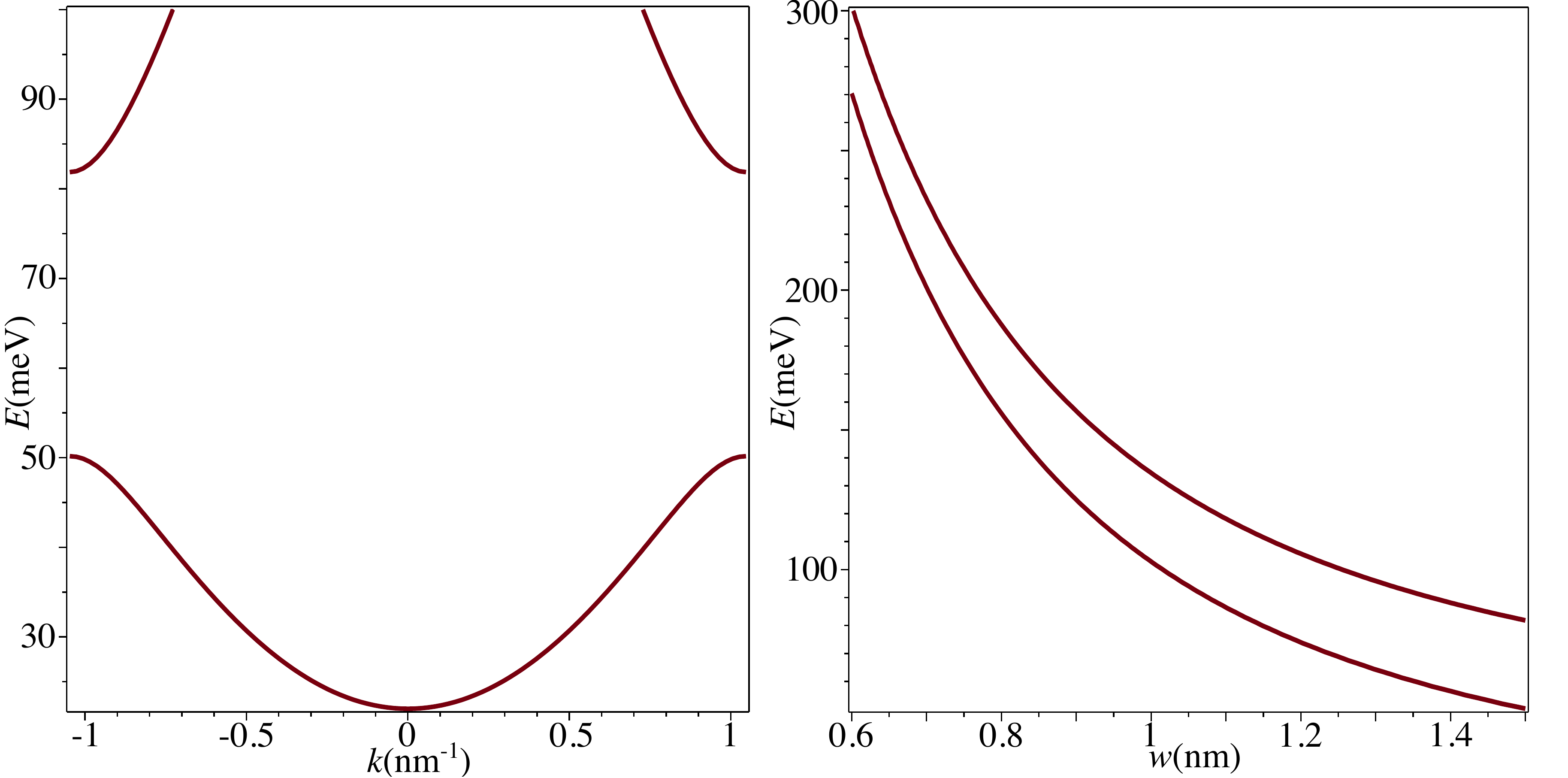}
\caption{\small Left panel: Band structure of a priodic superlattice with all regions with the same width given by $w_0=1.5$~nm. Right panel: The minigap between the first two minibands for the superlattice considered in the left panel as a function of the regions width.} \label{gap}
\end{figure}

These steps can be understood looking to Fig. \ref{gap}. In the left panel we have the first two minibands and the minigap between them of an infinite periodic superlattice, in which all regions have the same width given by $w_0$. In the right panel we can see that this minigap is shifted up when the width of the regions decreases. In the superlattice considered in this work, the widths of the regions are modulated between $w_0$ and 0. For small values of $\nu$, since the widths change only by a very small amount when $n$ increases, it is possible to observe the miniband and minigap structure locally. At this way, if the energy is inside of a minigap, the transmittance decays, while if it is in a miniband, the transmittance remains constant.

\begin{figure}[h]
\includegraphics[width=\linewidth]{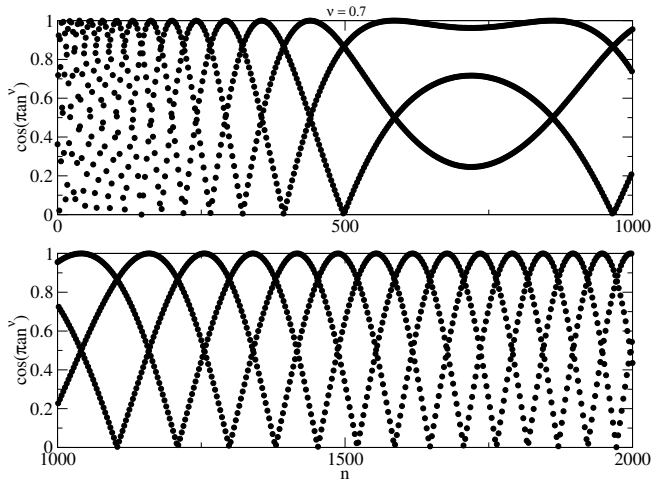}
\caption{\small The cosine that modulates the widths for $\nu=0.7$.} \label{V07}
\end{figure}

The dispersion relation for an one-dimensional periodic superlattice with all regions with the same width $w$ is given by
\begin{equation}
\cos 2kw=\cos k_1w \cos k_2w - \frac{k_1^2+k_2^2}{2k_1k_2}\sin k_1w \sin k_2w,
\end{equation}
where $k_1=\sqrt{2mE}/\hbar$, $k_2=\sqrt{2m(E-V)}/\hbar$ and $k$ is the Bloch wavenumber. For the widths between 0 and $w_0$, the energy of $40$~meV is always inside of a miniband. This explains the metallic phase observed in the top panel of Fig. \ref{TxNsmall}. The energy of $100$~meV is inside of a minigap for a superlattice with $w$ between $1.015-1.255$~nm. Dividing these values by $w_0$, we obtain that one may observe a decay in the transmittance when the absolute value of $\cos(\pi a n^{\nu})$ is between $0.67-0.83$. Looking to Fig. \ref{step}, we can confirm exactly this. So, this explains the steps in the transmittance observed in the bottom panel of Fig. \ref{TxNsmall}. As the value of $\nu$ increases, the period of oscillations of $|\cos(\pi a n^{\nu})|$ decreases. Due to this faster oscillation, it is not possible to identify the steps and plateaus in the transmittance for $\nu$ equal to 0.5 and 0.6.

In contrast to the cases with $\nu$ from 0.2 to 0.6, where the cosine that modulates the widths oscillates with $n$, for $\nu=0.7$ we can identify three oscillating curves, and the value of the cosine alternates among these three curves. It can be seen in Fig. \ref{V07}. For small values of $n$, we have such a distribution of the widths. As $n$ increases, we can identify clearly the three oscillating curves. However, only after $n=1000$ we can see a nearly periodic oscillation of these curves, with the period of oscillation decreasing as $n$ increases. It explains the initial decay in the transmittance for $\nu=0.7$ in the top panel of Fig. \ref{TxNsmall}. After $N=1000$, the transmittance remains constant, which characterizes a metallic phase for this case. 

\subsubsection{High values of $\nu$}

\begin{figure}[h]
\includegraphics[width=\linewidth]{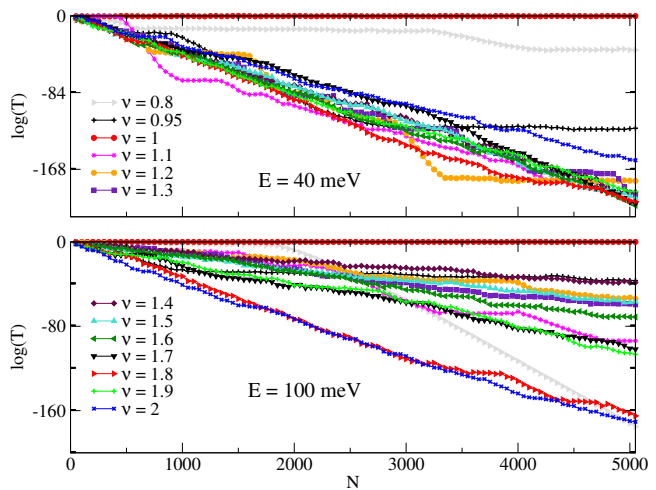}
\caption{\small The transmittance as a function of $N$ for different values of $\nu$.} \label{TxNhigh}
\end{figure}

In Fig. \ref{TxNhigh} we consider the transmittance as a function of $N$ for high values of $\nu$ from $0.8$ to $2$ for $E=40$~meV (top panel) and $E=100$~meV (bottom panel). We can see that the transmittance decays as the length of the system increases for $\nu \geq 1$, revealing an insulator phase. For $\nu$ slightly below unit, there is an initial decay of transmission before reaching a plateau. This will give an insulator character for systems with a small number of layers and a metallic one for very longer structures. In order to understand these results, let us first look to the case $\nu=0.8$.

\begin{figure}[h]
\includegraphics[width=\linewidth]{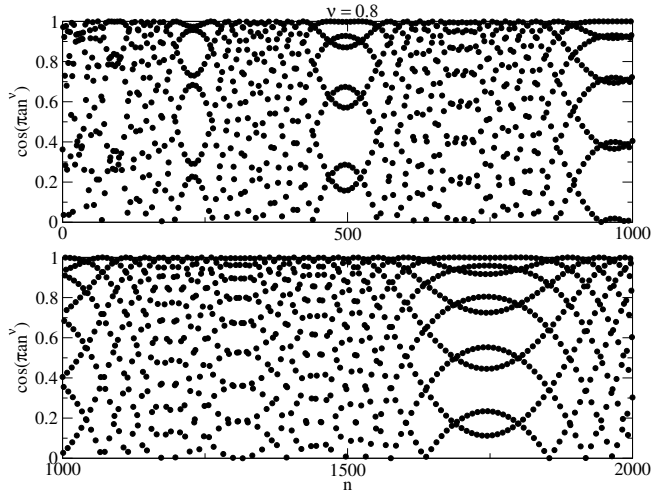}
\caption{\small The cosine that modulates the widths for $\nu=0.8$.} \label{V08}
\end{figure}

In Fig. \ref{V08} we consider the cosine that modulates the widths in the superlattice for the case $\nu=0.8$. We can see that, as in the case $\nu=0.7$, it is possible to identify some oscillating curves and the value of the cosine alternates among these curves. However, as $n$ increases, the number of oscillating curves also increases. For instance, around $n=250$, we can identify 6 curves. At $n=500$, we can see now 7 curves. This value increases to 8 around $n=1000$. Such behavior does not allow the appearance of the almost regular oscillations observed in the $0.7$ case, which gives rise to the insulator phase for this range of system sizes. Only for very large system sizes the quasi-periodic pattern develops and the transmission stabilizes at a small-valued plateau. 

\begin{figure}[h]
\includegraphics[width=\linewidth]{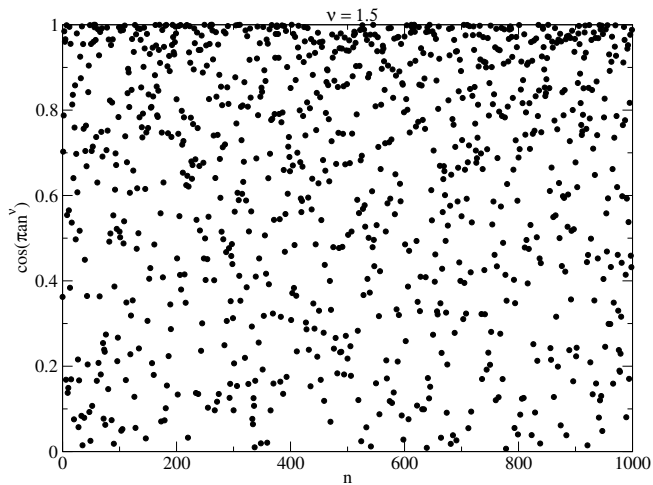}
\caption{\small The cosine that modulates the widths for $\nu=1.5$.} \label{V15}
\end{figure}

\begin{figure}[h]
\includegraphics[width=\linewidth]{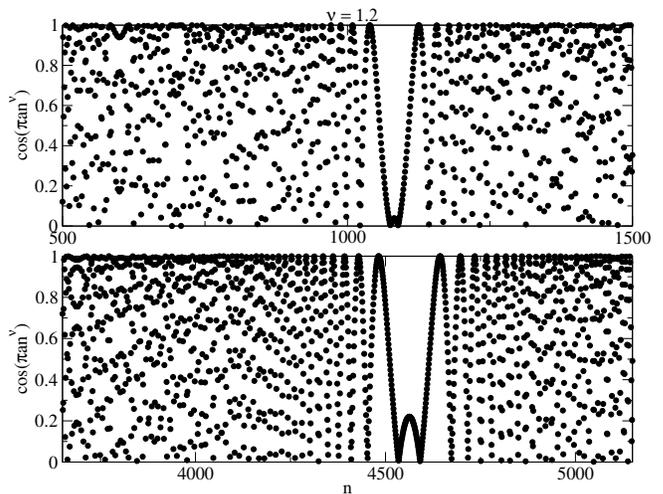}
\caption{\small The cosine that modulates the widths for $\nu=1.2$.} \label{V12}
\end{figure}

For higher values of $\nu$, we have a pseudo-random distribution of the widths of the regions in the superlattice, which justify the insulator phase observed in Fig. \ref{TxNhigh}. It can be seen, for instance, in Fig. \ref{V15}, where we considered the cosine that modulate the widths of the regions for the case of $\nu=1.5$.

\begin{figure}[h]
\includegraphics[width=\linewidth]{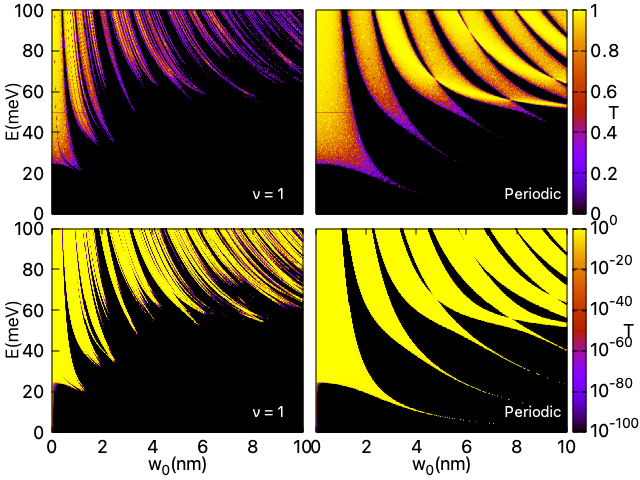}
\caption{\small (Left panel) T as a function of E and $w_0$ for $\nu=1$, N=5000 and $V=50$~meV. (Right panel) T as a function of E and $w_0$ for the periodic case with N=5000 and $V=50$~meV. Transmission is represented in linear scale on the top panels. In the bottom panels, it is represented on logarithmic scale for better visualization.} \label{V}
\end{figure}

We can also see in the top panel of Fig. \ref{TxNhigh} the presence of some plateaus in the transmittance for some values of $\nu$. For instance, when $\nu=1.2$, we have a plateau around $N=1000$ and $N=4500$. It happens because specifically in these regions there is a smooth modulation of the widths. The cosine that modulates the widths of the regions in the superlattice for $\nu=1.2$ can be seen in Fig. \ref{V12}. We can see that around $n=1000$ and $n=4500$ the widths of subsequent regions change only by a small amount, as in the case for small values of $\nu$, inducing the plateaus. Such plateaus are not seen for $E=100$~meV. It is a consequence of the minigap discussed previously for the case of small values of $\nu$. 

\subsubsection{The $\nu = 1$ case}

Let us now analyze the $\nu=1$ case. In contrast to the other values of $\nu$, when $\nu=1$ we have a metallic phase for both values of energy that we considered here, $40$~meV and $100$~meV. In order to understand this, we consider in the Fig. \ref{V} a contour plot of the transmittance as a function of the energy and the disorder strength $w_0$. We can identify such a miniband structure, where the minibands represent a metallic phase, while the minigaps an insulator phase. For some range of energy ($\approx$ 20 meV - 60 meV), we can see that for small values of $w_0$, the minibands are larger than the minigaps and as $w_0$ increases, the minigaps become larger than the minibands. Then, after certain value of $w_0$, there is only an insulator phase. 

\begin{figure}[h]
\includegraphics[width=\linewidth]{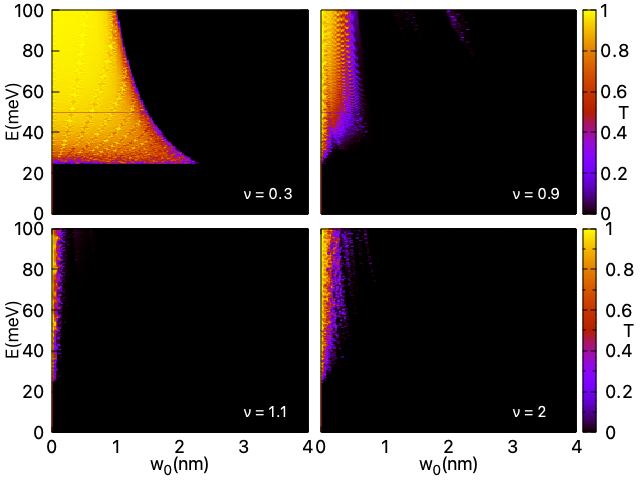}
\caption{\small T as a function of E and $w_0$ for the cases $\nu=0.3$, $0.9$, $1.1$ and $2$, respectively, with N=5000 and $V=50$~meV. } \label{twcases}
\end{figure}

For the sake of comparison, we also considered in Fig. \ref{V} the case of a periodic superlattice, which is obtained by replacing the cosine in Eq.(\ref{w}) by 1. We can see that this alternation between a metallic and insulator phases is reminiscent from the miniband and minigap structure of the periodic case. In order to leave clear the particular behaviour of the $\nu=1$ case, we also considered in Fig. \ref{twcases} the cases $\nu=0.3$, $0.9$, $1.1$ and $2$. It can be seen that there is a metallic phase only for small values of the disorder strength $w_0$, which means that when one increases $w_0$, there is only one metallic-insulator phase transition. Also, this metallic phase region decreases as we increase the value of $\nu$, revealing that for large values of $\nu$, a metallic-like behavior can only be reached for very small values of $w_0$. 

\begin{figure}[h]
\includegraphics[width=\linewidth]{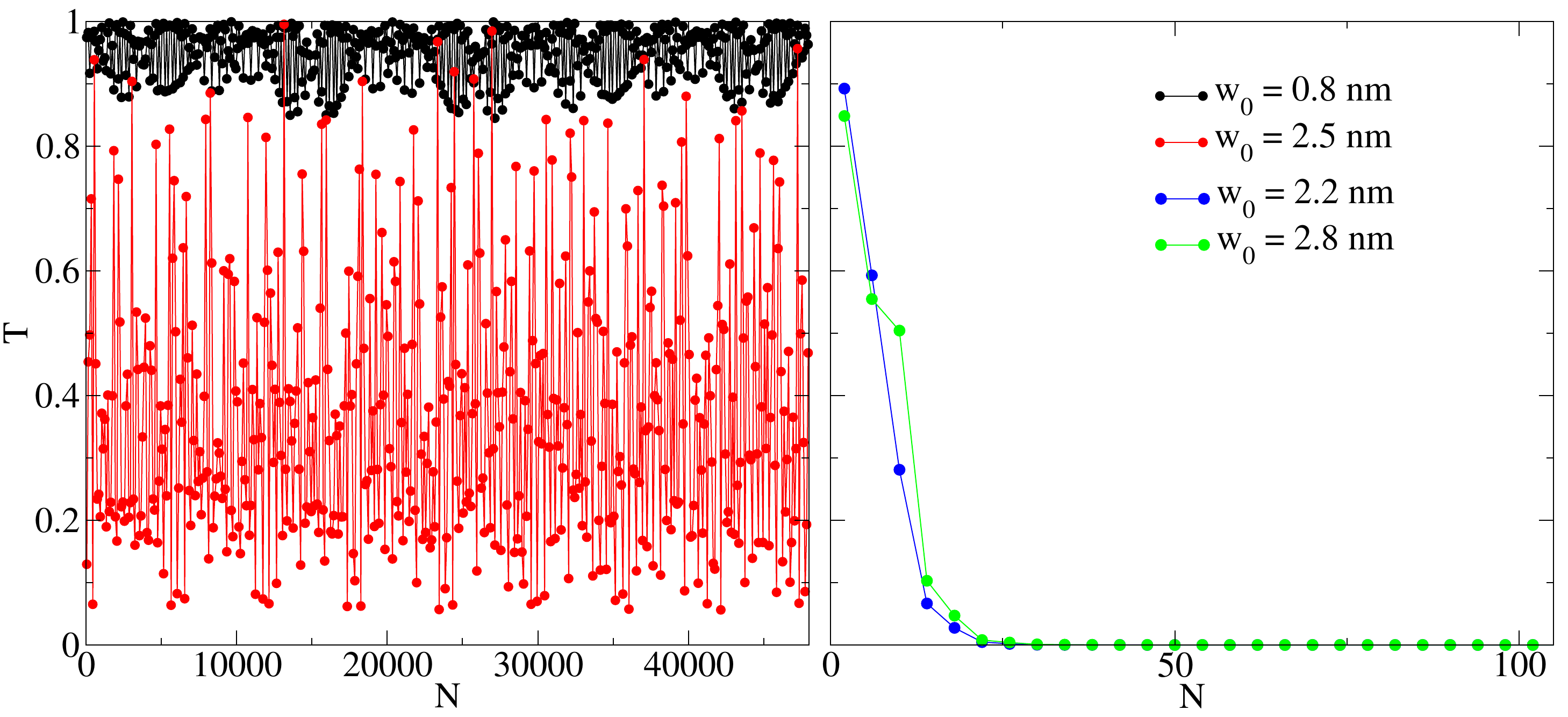}
\caption{\small T as a function of N for four different values of $w_0$ with $E=75$~meV and $\nu=1$.} \label{T}
\end{figure}

\begin{figure}[h]
\includegraphics[width=\linewidth]{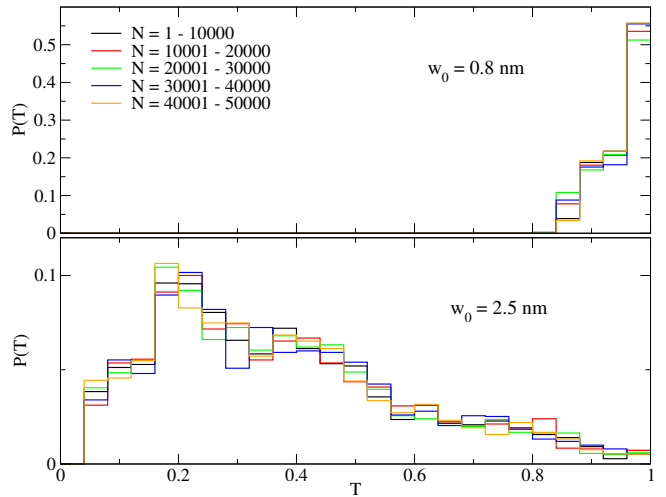}
\caption{\small The probability of the transmittance P(T). Left panel: $w_0=0.8$~nm. Right panel: $w_0=2.5$~nm.} \label{P}
\end{figure}

In Figs. \ref{T} and \ref{P} we analyze in details the metallic and insulator phases for $\nu = 1$. We consider in Fig. \ref{T} four different values of $w_0$, which reveals the alternation between a metallic and an insulator phase as we increase the disorder strength $w_0$.

The metallic phases are confirmed in Fig. \ref{P}, where we consider the probability of the transmittance P(T). We can see that, as we increase $N$, P(T) does not change. However, the highest pick of P(T) is observed for small values of T as we increase $w_0$. For instance, for $w_0=0.8$~nm, the transmittance is always between 0.85 and 1, while for $w_0=2.5$~nm the most likely values of T is around 0.2. 

\subsubsection{Metallic-insulator transitions}

\begin{figure}[h]
\includegraphics[width=\linewidth]{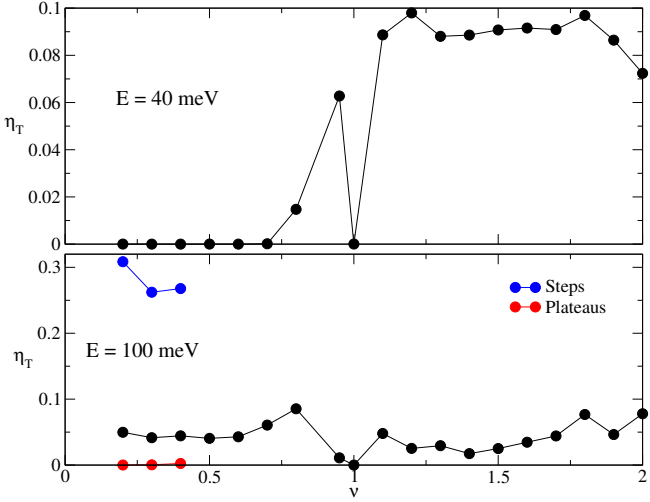}
\caption{\small The exponent $\eta_T$ as a function of $\nu$.} \label{etaT}
\end{figure}

In order to see the metallic-insulator transitions as a function of the parameter $\nu$, we write
\begin{equation}
T \propto \exp(-N\eta_T),
\end{equation}
where $\eta_T$ can be interpreted as the Lyapunov exponent. For $\eta_T = 0$ we have a metallic phase, while for $\eta_T \neq 0$ an insulator phase.

In Fig. \ref{etaT} we consider $\eta_T$ as a function of $\nu$. For $E=40$~meV, we can see clearly the metallic-insulator transition as we change the value of $\nu$. The raise of the Lyapunov exponent for $\nu$ just below unit is related to the fact that the initial exponential decay of the transmission survives for a long time before reaching a plateau. The Lyapunov exponent in this regime tends to decrease when it is measured considering structures with a larger number of layers. For $E=100$~meV, fitting the whole curve with an exponential function (black line), there is a metallic phase only for $\nu = 1$, which means that we do not have a metallic-insulator phase transition here. For $\nu=0.2, 0.3, 0.4$ we also obtained $\eta_T$ for the steps (blue line) and plateaus (red line) of the transmittance curve.  

\subsection{Oblique incidence case}

\begin{figure}[h]
\includegraphics[width=\linewidth]{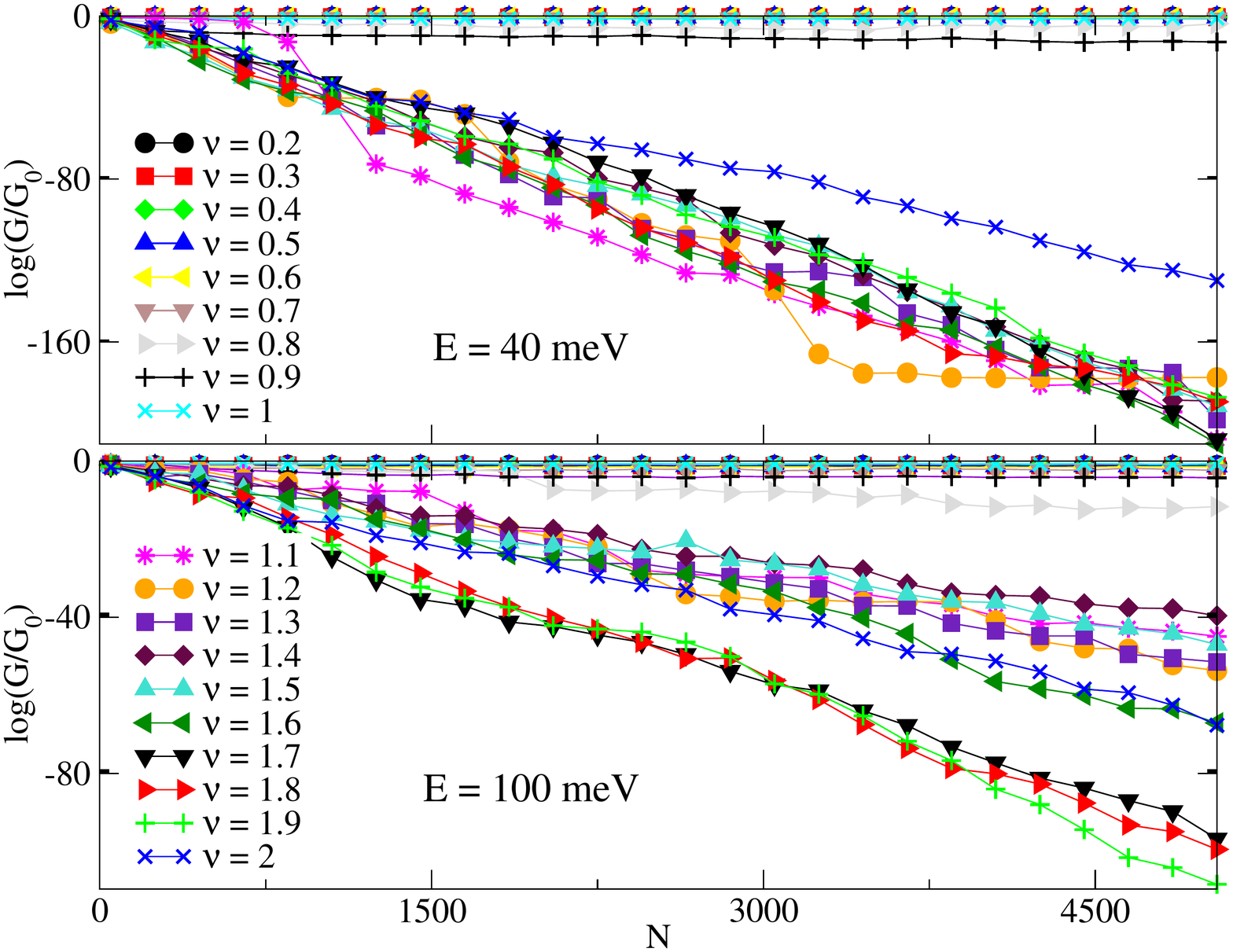}
\caption{\small The normalized conductance as a function of $N$ for different values of $\nu$.} \label{GxN}
\end{figure}

Let us now include the oblique incidence in our analysis, which means to consider a two-dimensional system. Here, we obtained the normalized conductance as a function of $N$ for different values of $\nu$, after integrating out over all incidence angles. It can be seen in Fig. \ref{GxN}, where we consider the $E=40$~meV (top panel) and $E=100$~meV (bottom panel). We can see that, depending on the value of $\nu$, the conductance remains constant or decays exponentially. 

\begin{figure}[h]
\includegraphics[width=0.8\linewidth]{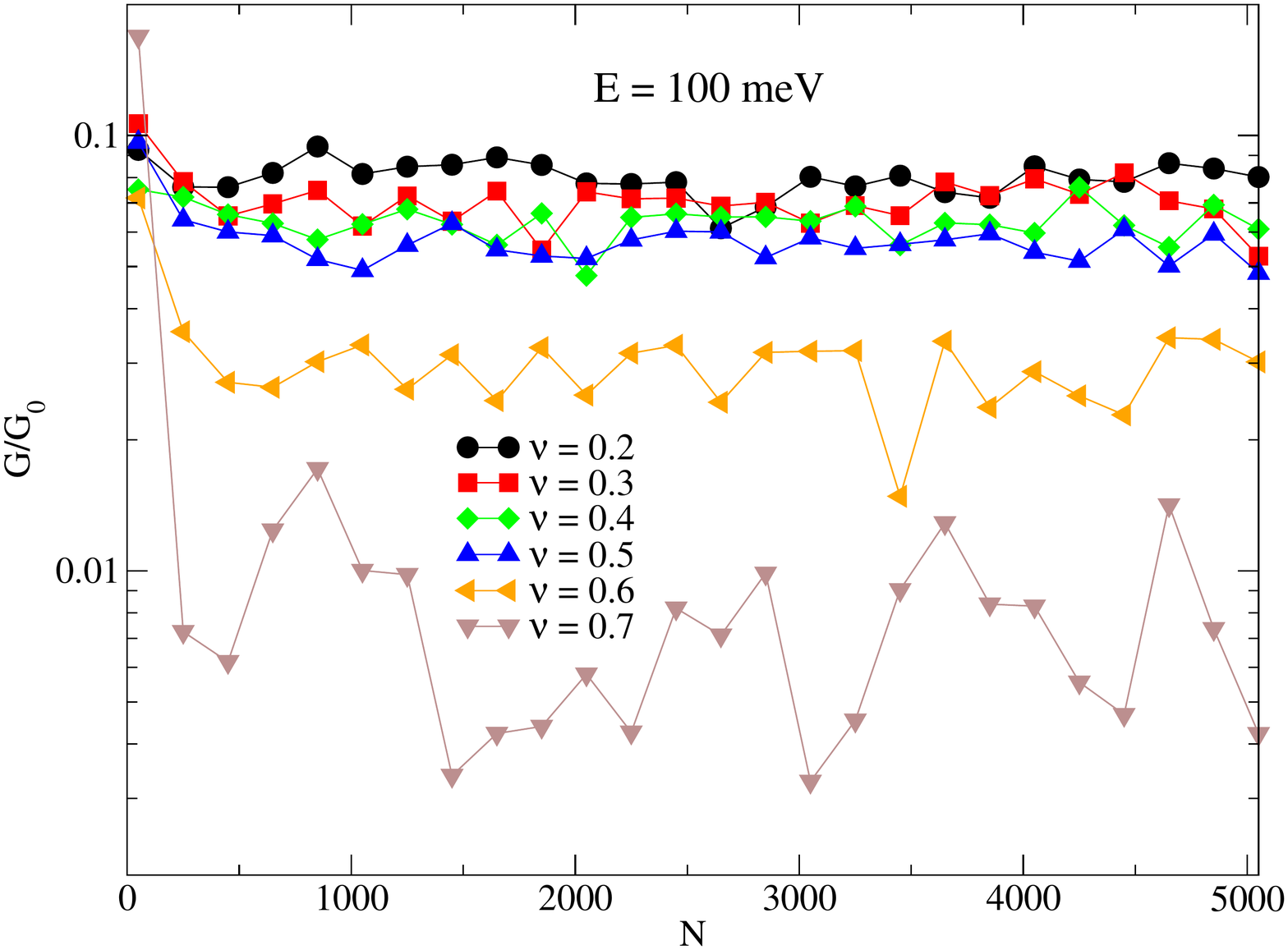}
\caption{\small The normalized conductance as a function of $N$ for different values of $\nu$.} \label{smallnu}
\end{figure}

As in the normal incidence case, for large values of $\nu$, the conductance decays exponentially due to the pseudo-random widths distribution. Also, for $\nu=1$, there is a metallic phase independent of the energy. However, in contrast to the normal incidence case, there is a metallic phase for small values of $\nu$ not only for $E=40$~meV, but also for $E=100$~meV. In fact, the conductance for $E=100$~meV initially decays exponentially and then becomes constant as we increase the value of $N$, which can be seen more clearly in Fig. \ref{smallnu}. 

\begin{figure}[h]
\includegraphics[width=0.8\linewidth]{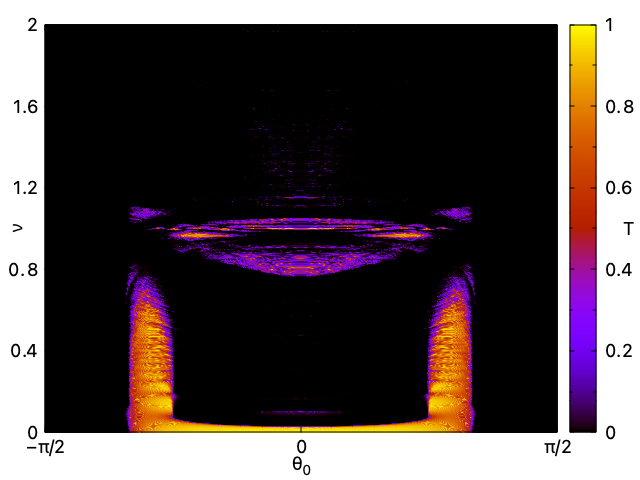}
\caption{\small Contour plot of the transmittance as a function of $\theta_0$ and $\nu$ for $E=100$~meV and $N=500$.} \label{nuth}
\end{figure}

In the 2D case, we have that $k_y=\frac{\sqrt{2mE}}{\hbar}\sin\theta_0$. Replacing it in Eqs. (\ref{k}) and (\ref{kp}), we have that in the empty regions 
\begin{equation}
k_x=\frac{\sqrt{2mE\cos^2\theta_0}}{\hbar},
\end{equation}
while in the barrier regions
\begin{equation}
k^{\prime}_x=\frac{\sqrt{2m(E\cos^2\theta_0-V)}}{\hbar}.
\end{equation}
We can see that the difference between the normal and oblique incidence cases is that for oblique incidence, we replace $E$ by $E\cos^2\theta_0$. It means that the oblique incidence case is equivalent of considering normal incidence with a reduced energy. In other words, integrating out the transmittance over all incidence angles to obtain the conductance is somehow equivalent of integrating the transmittance for the normal incidence over all energies from 0 to E. It can be seen in Fig. \ref{nuth}, where we consider the contour plot of the transmittance as a function of $\theta_0$ and $\nu$ for $E=100$~meV. We can see that the transmittance here is basically two copies of the transmittance in Fig. \ref{T1} for the energy from 0 to $100$~meV, which confirms what we have just discussed. 

Therefore, for small values of $\nu$, the transmittance decays exponentially for the energies that are inside of a minigap and remains constant for the other energies. At this way, there will be an exponential decay in the transmittance for some incidence angles ($\theta_G$) and a constant transmittance for others ($\theta_B$). So, the conductance initially decays as a consequence of the exponential decay of the transmission for the incidence angles $\theta_G$, and then, when there is only the contribution to the conductance from $\theta_B$, it remains constant in a metallic phase.  This explains the results showed in Fig. \ref{GxN}.

\begin{figure}[h]
\includegraphics[width=\linewidth]{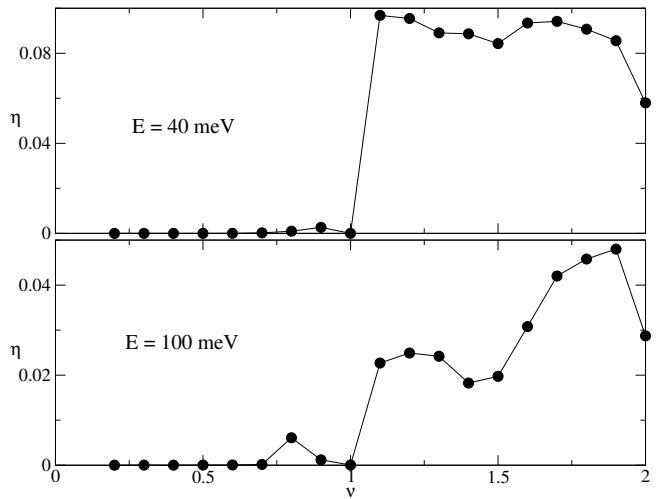}
\caption{\small The exponent $\eta$ as a function of $\nu$.} \label{etaG}
\end{figure}

We can write 
\begin{equation}
\frac{G}{G_0} \propto \exp(-N\eta),
\end{equation}
where $\eta$ is the Lyapunov exponent for the conductance. So, it is possible to plot $\eta$ as a function of $\nu$, which reveals clearly the metallic and insulator phases. Such plot can be seen in Fig. \ref{etaG}. So, for the 2D case, there is a metallic-insulator phase transition for both energies considered here. The small raise below $\nu=1$ is a finite-size effect resulting from the quasi-periodic nature of the widths distribution. It is consistent with the behavior found for the Lyapunov exponent computed from the own transmission. Here, the finite-size effect is less pronounced and also tends to disappear when much larger structures are considered.

\section{Conclusion}

In this work we have investigated the transport properties of non-relativistic quantum particles in an one-dimensional incommensurate superlattice that follows an extended Harper model. In our case, this model modulates the widths of each region in the superlattice. We have found novel metallic-insulator phase transitions. These transitions occur when we change the disorder strength $w_0$ and also the value of the exponent $\nu$. For the normal incidence case, there will be a phase transition as a function of $\nu$ only for a specific range of energy. This range of energy increases considerably when we look to the conductance, which is obtained integrating out the transmittance for all incidence angles. On the other hand, when one increases the disorder strength $w_0$, there is a metallic-insulator phase transition for all values of $\nu$. As we increase the value of $\nu$, this phase transition occurs at a lower value of $w_0$. However, for the special case of $\nu=1$, instead of a single metallic-insulator transition, there is an alternation between a metallic and an insulator phase, which is a particular behaviour of the system that we considered here. All the results obtained here contribute to a better understanding of quasiperiodic and incommensurate systems. 

\begin{acknowledgments}
JRFL thanks Igor Gornyi for useful and valuable discussions during his stay at the Karlsruhe Institute of Technology. This work was partially supported by Brazilian agency Conselho Nacional de Desenvolvimento Cient\'ifico (CNPq). 
\end{acknowledgments}


\begin{thebibliography}{99}
\bibitem{mermin} N. W. Ashcroft and N. D. Mermin, Solid State Physics (Holt, Rinehart and Winston, New York, 1976).

\bibitem{oxford} S. H. Simon, The Oxford Solid State Basics (Oxford University Press, Oxford, 2013). 

\bibitem{anderson} P.W. Anderson, Phys. Rev. \textbf{109}, 1492 (1958).

\bibitem{lee} P. A. Lee and T. V. Ramakrishnan, Rev. Mod. Phys. \textbf{57}, 287 (1985).

\bibitem{dimer1} D.H. Dunlap, H.-L. Wu and P.W. Phillips, Phys. Rev. Lett. \textbf{65}, 88 (1990).

\bibitem{dimer2} P.W. Phillips and H.-L. Wu, Science \textbf{252}, 1805 (1991). 

\bibitem{lyra} F.A.B.F. de Moura and M.L. Lyra, Phys. Rev. Lett. \textbf{81}, 3735 (1998).

\bibitem{izrailev} F.M. Izrailev and A.A. Krokhin, Phys. Rev. Lett. \textbf{82}, (1999).

\bibitem{sokoloff} J. B. Sokoloff, Phys. Rep. \textbf{126}, 189 (1985).

\bibitem{aubry} S. Aubry and G. André, Ann. Isr. Phys. Soc. \textbf{3}, 133 (1980).

\bibitem{fishman} M. Griniasty and S. Fishman, Phys. Rev. Lett. \textbf{60}, 1334 (1988).

\bibitem{sarma1} S. Das Sarma, S. He and X. C. Xie, Phys. Rev. Lett. \textbf{61}, 2144 (1989).

\bibitem{sarma2} S. Das Sarma, S. He and X. C. Xie, Phys. Rev. B \textbf{41}, 5544 (1990).

\bibitem{thouless} D. J. Thouless, Phys. Rev. Lett. \textbf{61}, 2141
(1988).

\bibitem{luschen} H. P. L\"uschen, P. Bordia, S. Scherg, F. Alet, E. Altman, U. Schneider, and I. Bloch, Phys. Rev. Lett. \textbf{119}, 260401 (2017).

\bibitem{schreiber} M. Schreiber, S.S. Hodgman, P. Bordia, H.P. L\"uschen,
M.H. Fischer, R. Vosk, E. Altman, U. Schneider, and I. Bloch, Science \textbf{349}, 842 (2015).

\bibitem{roati} G. Roati, C. D’Errico, L. Fallani, M. Fattori, C. Fort, M. Zaccanti, G. Modugno, M. Modugno, and M. Inguscio, Nature (London) \textbf{453}, 895 (2008).

\bibitem{eilmes} A. Eilmes, R. A. Romer and M. Schreiber, Eur. Phys. J. B \textbf{23}, 229 (2001).

\bibitem{moura} F. A. B. F. de Moura, M. L. Lyra, F. Dominguez-Adame, and V. A. Malyshev, Phys. Rev. B \textbf{71}, 104303 (2005).

\bibitem{peixoto} A. S. Peixoto, W. S. Dias, M. L.  Lyra, and F. A. B. F. de Moura, Physica A \textbf{395}, 22 (2014).

\bibitem{morales} L. Morales-Molina, E. Doerner, C. Danieli, and S. Flach, Phys. Rev. A \textbf{90}, 043630 (2014).

\bibitem{ray} S. Ray, M. Pandey, A. Ghosh, and S. Sinha, New J. Phys. \textbf{18}, 013013 (2015).

\bibitem{malla} R. K. Malla and M. E. Raikh, Phys. Rev. B \textbf{97}, 214209 (2018).

\bibitem{patra} M. Patra and S. K. Maiti, J. Mag. Mag. Mater. \textbf{ 484}, 408 (2019).

\bibitem{saha} M. Saha and S. K. Maiti, J. Appl. Phys. D \textbf{52}, 465304 (2019).

\bibitem{settino} J. Settino, N. W.  Talarico, F. Cosco, F. Plastina, S. Maniscalco, and N. Lo Gullo, Phys. Rev. B \textbf{101}, 144303 (2020).

\bibitem{cookmeyer} T. Cookmeyer, J. Motruk and J. E. Moore, Phys. Rev. B \textbf{101}, 174203 (2020).


\bibitem{sarkar} S. Sarkar and S. K.  Maiti, J. Phys.: Condens. Matt. \textbf{32}, 505301 (2020).

\bibitem{sajid} S. Sajid and A. Chakrabarti, Phys. Rev. B \textbf{102}, 134401 (2020).

\bibitem{yoo} Y. C. Yoo, J. H. Lee and B. Swingle, Phys. Rev. B \textbf{102}, 195142 (2020).

\bibitem{roy} N. Roy and A. Sharma, Phys. Rev. B \textbf{100}, 195143 (2019).

\bibitem{jiang} H. Jiang, L. J. Lang, C. Yang, S. L. Zhu, and S. Chen, Phys. Rev. B \textbf{100}, 054301 (2019).

\bibitem{gong} L. Y. Gong, H. Zhu, S. M. Zhao, W. W. Cheng, and Y. B. Sheng, Phys. Lett. A \textbf{376}, 3026 (2012).

\bibitem{messias} D. Messias, C. V. C. Mendes, G. M. A. Almeida, M. L. Lyra, and F. A. B. F.  de Moura, J. Mag. Mag. Mater. \textbf{505}, 166730 (2020).

\bibitem{nascimento} E. M. Nascimento, F. A. B. F.  de Moura and M. L.  Lyra, Photonic and Nanostructures \textbf{7}, 101 (2009).

\bibitem{soto} Luis L. Sánchez-Soto, Juan J. Monzón, Alberto G. Barriuso and José F. Cariñena, Physics Reports  \textbf{513}, 191 (2012).

\bibitem{conduc} M. Ramezani Masir, P. Vasilopoulos, and F. M. Peeters, Phys. Rev. B  \textbf{82}, 115417 (2010).


\end{thebibliography}
\end{document}